\documentclass[11pt]{article}
\usepackage{hyphenat}
\usepackage[final]{acl}

\usepackage{times}
\usepackage{latexsym}

\usepackage[T1]{fontenc}
\usepackage[utf8]{inputenc}
\usepackage{microtype}
\usepackage{inconsolata}
\usepackage{graphicx}

\usepackage{times}
\usepackage{latexsym}
\usepackage[T1]{fontenc}
\usepackage{amsmath}
\usepackage{amssymb}
\usepackage{tcolorbox}
\usepackage{booktabs}
\usepackage[table]{xcolor}
\usepackage{makecell}
\usepackage{hhline}
\usepackage{array}
\usepackage{colortbl}
\usepackage{tabularx}
\usepackage{threeparttable}
\usepackage{multirow}

\usepackage[utf8]{inputenc}
\usepackage{microtype}
\usepackage{inconsolata}
\usepackage{graphicx}
\usepackage{xcolor}
\usepackage{enumitem}
\usepackage{algorithm}
\usepackage{algpseudocode}
\usepackage{subcaption}
\usepackage{listings}

\definecolor{headblue}{RGB}{210,225,245}
\definecolor{rowgray}{RGB}{245,247,250}
\newcommand{\cc}[1]{\cellcolor{rowgray}{#1}}
\newcommand{\ccd}[1]{\cellcolor{headblue}\textbf{#1}}
\newcolumntype{Y}{>{\raggedright\arraybackslash}X}

\lstdefinestyle{promptblock}{
  basicstyle=\ttfamily\scriptsize,
  breaklines=true,
  breakatwhitespace=true,
  frame=single,
  rulecolor=\color{gray!50},
  backgroundcolor=\color{gray!4},
  xleftmargin=0.8em,
  xrightmargin=0.5em,
  framexleftmargin=0.5em,
  aboveskip=0.1em,
  belowskip=0.1em,
  columns=fullflexible,
  keepspaces=true,
  showstringspaces=false,
  lineskip=-0.9pt,
  captionpos=b,
  fontadjust=true,
}

\title{MACAA: Belief-Revision Multi-Agent Reasoning for Code Authorship Verification}

\author{
Jingwei Ye$^{1}$,
Zhi Wang$^{1}$\thanks{Corresponding author.},
Xin Li$^{1}$,
Cong Gao$^{1}$,
Chenbin Su$^{1}$,
Jieshuai Yang$^{1}$, \\
\textbf{Jianfei Tang$^{1}$, Ge Chu$^{2}$} \\
$^{1}$College of Cryptology and Cyber Science, Nankai University, China \\
$^{2}$Runjian Co., Ltd., China \\
}

\begin{document}
\maketitle

\begin{abstract}
Code authorship attribution (CAA) supports software forensics, plagiarism detection, and intellectual property protection. However, existing supervised CAA approaches suffer from scarce training data and closed-world assumptions: they require sufficient labeled code from fixed candidate-author sets, making training difficult in low-data cases and predictions unreliable for open-world test pairs with unseen samples, or heterogeneous code pairs. Large language models remove task-specific training, but direct prompting depends on costly expert-designed prompts, can hallucinate over complex heterogeneous code pairs, and rarely yields auditable evidence traces. We propose MACAA, a belief-revision-based multi-agent framework for training-free code authorship verification. MACAA comprises a Coordinator and four Expert Agents analyzing layout, lexical, syntactic, and programming-pattern evidence. The Coordinator gathers expert signals for expansion, discounts unreliable evidence through contraction, and resolves conflicts through revision to preserve belief consistency, replacing direct LLM judgment with auditable hypothesis refinement. MACAA achieves 89.15\% F1 on same-language benchmarks and 80.00\% on mixed cross-language pairs, outperforming the baselines overall in both same-language and cross-language evaluations.\footnote{Our code is released at \url{https://github.com/2845731/Multi-Agent-Reasoning-for-Code-Authorship}.}

\end{abstract}

\section{Introduction}
\label{sec:introduction}

\begin{figure*}[t]
    \centering
    \includegraphics[width=0.9\textwidth]{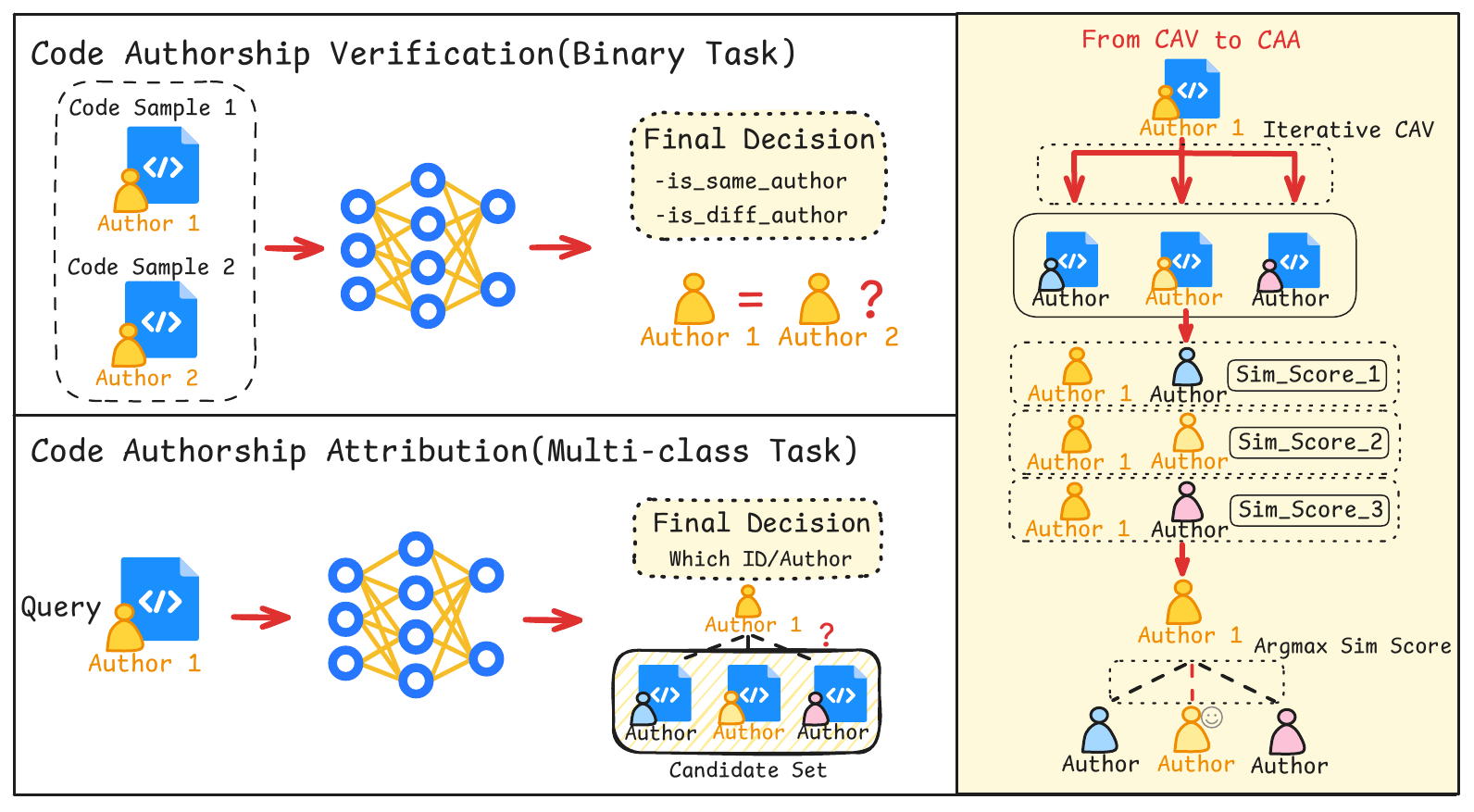}
    \caption{From Verification to Attribution in Code Authorship}
    \label{fig:cav_to_caa}
\end{figure*}

Code authorship attribution (CAA) supports software forensics, plagiarism detection, and intellectual property protection. CAA includes several related tasks, such as identification, verification, clustering, evolution tracking, and author profiling. This paper focuses on \emph{code authorship verification} (CAV), the binary task of deciding whether two code samples were written by the same author. Figure~\ref{fig:cav_to_caa} illustrates this relation: the upper part shows pairwise CAV, the lower part shows multi-class CAA over a query and candidate set, and the right part shows how repeated CAV produces similarity scores for candidate ranking. We study CAV because it is the atomic operation underlying CAA and is better suited to forensic settings than closed-world identification. In practice, the true author may be absent from any fixed candidate pool, and only limited code evidence may be available. Verification is therefore naturally an open-world problem rather than a closed-set classification task \cite{kalgutkar2019code, choi2025can}.


Supervised CAA methods face two main limitations in such settings: scarce effective training data and closed-world generalization. They require code samples paired with reliable author labels, but practical investigations may provide too few samples or samples without usable labels. Even when labeled data exist, supervised models usually learn from a fixed set of candidate authors and assume that test samples follow the training distribution. When deployed in open-world settings, where test samples may involve unseen authors, projects, or language pairs, the learned decision boundary can become unreliable.

Cross-language verification further weakens surface-level authorship cues. In heterogeneous code pairs, the same author may produce very different surface forms because programming languages impose different syntax, type systems, library ecosystems, and idiomatic patterns \cite{guo2024enhancing, kalgutkar2019code}. Therefore, surface differences should not be treated as direct evidence of different authors. A reliable verifier must distinguish author-induced stylistic habits from language-induced variation and down-weight signals caused mainly by the programming language itself.

Large language models (LLMs) provide a training-free alternative for CAA, but existing LLM-based reasoning schemes remain insufficient. Direct prompting places most task knowledge inside the prompt: strong performance often requires expert-designed instructions that specify which stylistic cues to inspect and how to weigh them. Chain-of-thought prompting can expose intermediate reasoning, but on complex and heterogeneous code pairs, long reasoning traces may become unfaithful or hallucinated, causing the final decision to rest on weak evidence \cite{turpin2023language}. ReAct-style agents improve interaction by interleaving reasoning and actions \cite{yao2023react}, yet a single agent still lacks independent evidence checks across authorship dimensions. Pipeline-style multi-agent systems decompose the task, but errors can propagate from earlier agents to later stages without explicit belief correction. Debate-based systems encourage agents to challenge one another \cite{du2024improving, chan2023chateval}, but consensus does not guarantee correctness and may favor agreement over conflict resolution \cite{wynn2025talk}. These limitations motivate a framework that can decompose evidence, revise unreliable beliefs, and record an auditable decision path.

Both supervised methods and direct LLM prompting share a deeper structural limitation: they treat authorship verification as a single-pass decision. This is fragile when evidence is mixed. A reliable verifier should instead maintain a working hypothesis, update it as evidence accumulates, weaken unreliable signals, resolve conflicts, and record why each change was made. Belief revision theory, formalized by \citet{alchourron1985logic} through the AGM paradigm, provides a principled framework for this process. Its core operations---\textbf{expansion}, \textbf{contraction}, and \textbf{revision}---support adding new evidence, weakening unreliable beliefs, and restoring consistency under conflicting information \cite{aravanis2023collective}.

Motivated by these limitations and the belief revision principle, we propose \textbf{MACAA} (\textbf{M}ulti-\textbf{A}gent \textbf{C}ode \textbf{A}uthorship \textbf{A}ttribution). The Coordinator Agent acts as the central controller, maintaining a preliminary authorship hypothesis and orchestrating the expert analyses and belief-revision workflow. The Expert Agents analyze layout, lexical, syntactic, and programming-pattern evidence. Their analyses support expansion by adding dimension-specific evidence to the shared working memory. Cross-language calibration and agent rechecking support contraction by reducing unreliable, language-induced, or inconsistent signals. Structured debate further reconciles conflicts among Expert Agents. Finally, the Synthesize--Reflect--Finalize stage performs revision: it decides whether to preserve or update the preliminary hypothesis and produces the final decision. The stored analyses and belief updates serve as contextual memory and provide a traceable evidence path. 

Our contributions are threefold:
\begin{enumerate}[leftmargin=*]
    \item \textbf{Training-Free Open-World Verification:} We formulate code authorship verification as a training-free open-world task that compares code pairs without labeled training data or closed-world candidate-author assumptions, better reflecting real-world forensic scenarios with limited code evidence and no fixed author pool.

    \item \textbf{AGM-Inspired Belief Revision Multi-Agent Architecture:} We introduce MACAA, which draws on the operations of the AGM paradigm---expansion, contraction, and revision---through a Coordinator Agent and four Expert Agents that analyze layout, lexical, syntactic, and programming-pattern evidence.

    \item \textbf{Explicit Evidence Tracing for Auditable Reasoning:} MACAA records the belief revision path, including which evidence dimensions are expanded, weakened, or revised, enabling investigators to inspect the feature-level basis of each decision.
\end{enumerate}

\section{Related Work}
\label{sec:related-work}

\textbf{Code Authorship Analysis.} First early methods relied on manual feature engineering extracting lexical patterns, layout conventions, and syntactic statistics \cite{burrows2007source, caliskan2015anonymizing}, while subsequent neural approaches employed recurrent, convolutional, or transformer architectures to capture stylistic regularities \cite{alsulami2017source, ullah2019source}. These methods assume closed-world settings requiring labeled training data for fixed author sets. Second, \citet{quiring2019misleading} demonstrated vulnerability to adversarial transformations, prompting defenses like Forsee leveraging expert feature knowledge for robust attribution \cite{guo2024enhancing}. Third, Cross-language analysis has been addressed through language-oblivious features \cite{abuhamad2018large} and platform-independent approaches \cite{abazari2022language}, though these still require author-specific supervision. Recent work by \citet{ou2022scs} argues verification is more practical than closed-set identification for forensics, though their approach remains trained. LLMs offer zero-shot reasoning without task-specific training \cite{choi2025can}, but direct prompting produces non-iterative decisions lacking transparency.

\textbf{Multi-Agent Systems.} 
Prompting methods improve LLM reasoning by eliciting intermediate steps ~\cite{wei2022chain,kojima2022large}. Single-agent frameworks such as ReAct and Toolformer combine reasoning with tool use ~\cite{yao2023react,schick2023toolformer}, while multi-agent frameworks enable task decomposition and specialized collaboration through communication ~\cite{li2023camel,wu2024autogen,hong2024metagpt}. However, these frameworks mainly emphasize coordination rather than explicitly correcting unreliable or conflicting evidence. This limitation is important for code authorship verification, where shared templates, language conventions, or project-specific styles may be mistaken for genuine authorial habits.


\textbf{Belief Revision.} 
The AGM framework \cite{alchourron1985logic} formalizes how beliefs should change when new information arrives, while preserving consistency and minimizing unnecessary revision. Subsequent work extends belief revision to collective settings \cite{aravanis2023collective} and develops semantic accounts of belief update and AGM-style revision \cite{bonanno2025kripke}. These studies provide a useful foundation for reasoning with mixed or conflicting evidence. MACAA builds on this view by modeling authorship verification as iterative evidence accumulation, reliability adjustment, and hypothesis refinement.

\textbf{Positioning.} MACAA differs from prior code authorship analysis by framing verification as \textbf{iterative belief revision} rather than static feature matching; differs from existing multi-agent frameworks by \textbf{drawing on the AGM paradigm} for principled belief update; and differs from prior belief revision applications by \textbf{tailoring to heterogeneous code verification} with cross-language calibration and auditable reasoning.

\begin{figure*}[htbp]
\centering
\includegraphics[width=1\textwidth]{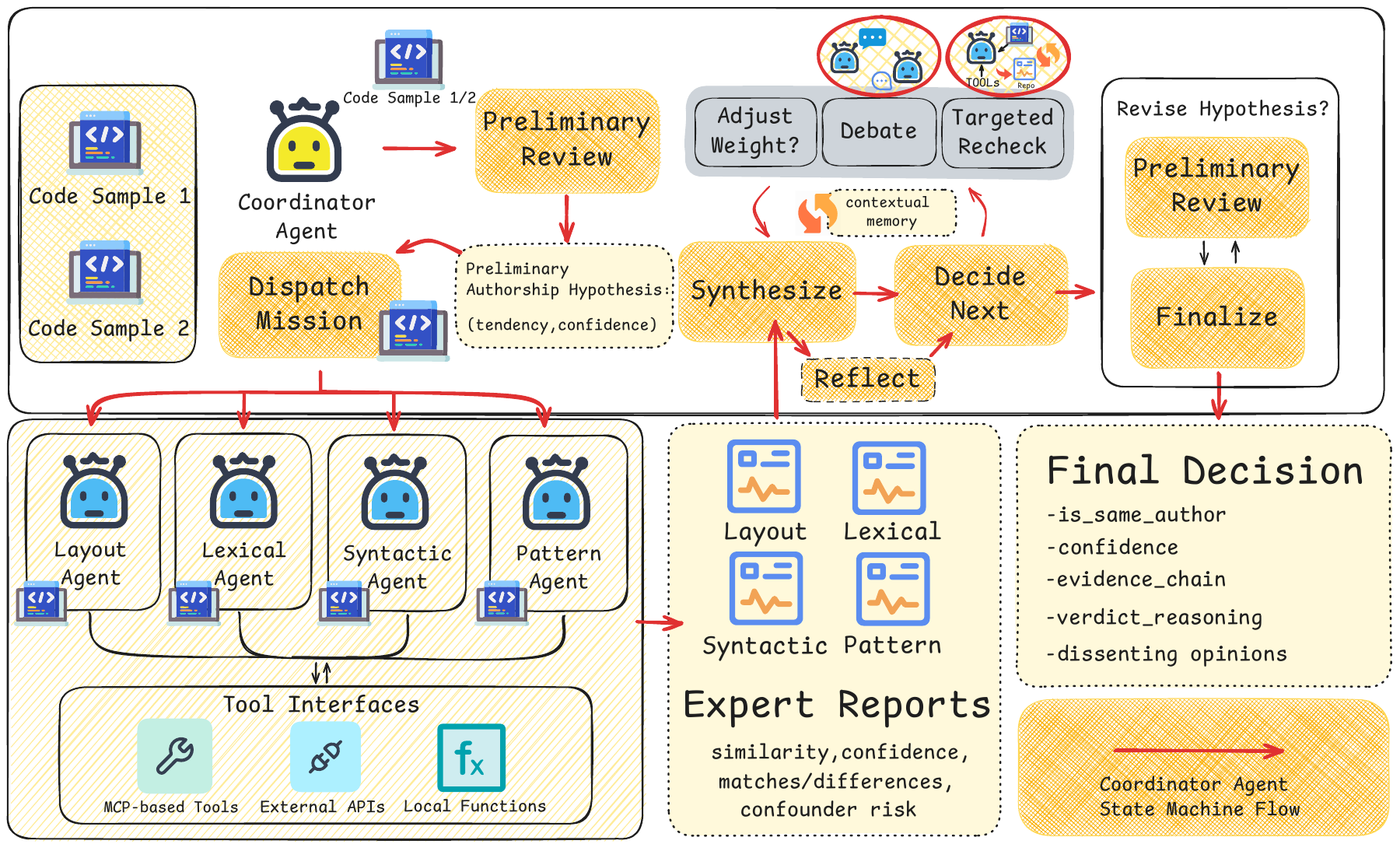}
\caption{MACAA overview with Coordinator Agent state-machine flow for expert evidence analysis, belief revision, and auditable authorship decisions.}
\label{fig:MAS_framework}
\end{figure*}

\section{Methodology}
\label{sec:methodology}

In this section, we present the details of the MACAA framework. As illustrated in Figure~\ref{fig:MAS_framework}, MACAA comprises five core components: a Coordinator Agent that manages the working belief state and orchestrates the verification workflow, and four Expert Agents that analyze complementary evidence dimensions---layout, lexical, syntactic, and programming patterns. Detailed prompt designs for all agents are provided in Appendix~\ref{sec:appendix-agent-prompts}.

\subsection{Belief Revision Foundation}
\label{sec:belief-revision}


Our design is inspired by the AGM theory of belief revision \cite{alchourron1985logic}, which formalizes how rational agents update beliefs upon receiving new evidence while preserving consistency and minimizing unnecessary changes. We adopt the operational vocabulary of expansion, contraction, and revision as a high-level organizational principle rather than claiming formal adherence to every AGM postulate. The three AGM operations serve as design labels for MACAA's architectural components as follows:

\vspace{2pt}
\noindent\textbf{Expansion} ($K + \phi$)---Adding consistent new beliefs to the knowledge base $K$ when evidence $\phi$ arrives. In MACAA, this corresponds to the \textbf{Expert Agents} gathering evidence from their respective dimensions and contributing observations to the working belief state.

\noindent\textbf{Contraction} ($K \div \phi$)---weakens or removes beliefs from $K$ when new evidence reveals unreliable or conflicting assumptions, while preserving as much useful information as possible. In MACAA, contraction is operationalized by the \textsc{Adjust Weight} step. The Coordinator adjusts the reliability weights of evidence dimensions based on three sources: cross-language calibration, targeted rechecking, and structured debate. Cross-language calibration down-weights signals likely induced by language-specific syntax, libraries, or idioms. Targeted rechecking revisits uncertain dimensions to verify weak evidence. Structured debate compares conflicting expert claims and exposes which dimension provides stronger support. These operations do not finalize the authorship decision; instead, they calibrate dimension-level evidence before the later revision stage decides whether to preserve or update the preliminary authorship hypothesis.

\noindent\textbf{Revision} ($K * \phi$)---incorporates calibrated evidence into the current belief state while preserving consistency. In MACAA, the Synthesize--Reflect--Finalize stage performs revision: after debate reconciles conflicting expert evidence and contraction weakens unreliable signals, the Coordinator preserves or updates the preliminary authorship hypothesis using the remaining credible evidence.

The principle of minimal change guides how MACAA updates beliefs under conflict. Instead of discarding the preliminary authorship hypothesis, MACAA makes local adjustments to evidence reliability: it weakens only the dimensions whose evidence is unreliable, inconsistent, or weakly supported, while preserving the remaining credible evidence in the working belief state. Thus, the framework revises the parts of the belief state that cause conflict without unnecessary changes to the overall decision process.

\subsection{Problem Formulation}
\label{sec:problem-formulation}

Given two code samples $x_1$ and $x_2$, potentially written in distinct programming languages, MACAA predicts whether they originate from the same author. The final decision includes: (1) a binary label $\texttt{is\_same\_author} \in \{\text{true}, \text{false}\}$; (2) a confidence score $c \in [0,1]$; (3) an evidence chain $\mathcal{E}=\{e_1,\ldots,e_k\}$ documenting the belief-revision trajectory; (4) verdict reasoning that summarizes the decisive evidence; and (5) dissenting opinions that record unresolved conflicts, weak signals, or reliability concerns for auditable reasoning.

\subsection{Coordinator Agent}
\label{sec:coordinator}

The Coordinator Agent implements the state-machine flow shown by the red arrows
in Figure~\ref{fig:MAS_framework}.
Given the input pair $(x_1, x_2)$ from \S\ref{sec:problem-formulation}, it acts
as the central controller: it maintains the preliminary authorship hypothesis,
dispatches Expert Agents, updates the shared \texttt{CaseState}, and coordinates
expansion, contraction, and revision.
The \texttt{CaseState} stores \texttt{expert\_reports},
\texttt{synthesis\_history}, and \texttt{reflection\_history} as contextual
memory for later reasoning and provenance reconstruction.

\textbf{State Machine Design.} The Coordinator proceeds as follows.
\textsc{Preliminary Review} performs an initial comparison of $x_1$ and $x_2$
and initializes $K_0$ with a tendency and a preliminary confidence score
$c_0 \in [0,1]$.
\textsc{Dispatch Mission} invokes the Layout, Lexical, Syntactic, and Pattern
Agents on $(x_1, x_2)$.
Their reports support expansion ($K+\phi$) by contributing
dimension-specific evidence items $e_i$ to the growing evidence chain
$\mathcal{E}=\{e_1,\ldots,e_k\}$, each including similarity, confidence,
matches or differences, and confounder risks.
\textsc{Synthesize} aggregates these reports under the current weights into a
structured evidence summary, rather than directly finalizing the belief state.
\textsc{Reflect} compares the current synthesis with contextual memory; it is
skipped in the first pass because no prior synthesis or reflection history
exists.
\textsc{Decide Next} first determines whether the current evidence is sufficient
for \textsc{Finalize}, or whether the maximum iteration budget has been reached.
Otherwise, it selects \textsc{Targeted Recheck} for uncertain dimensions or
\textsc{Debate} for conflicting expert judgments.
The resulting recheck or debate context is stored in \texttt{CaseState} and used
by the Coordinator to \textsc{Adjust Weight}, implementing contraction
($K\div\phi$) by strengthening credible dimensions and weakening unreliable,
language-induced, or inconsistent evidence.
Once the loop exits, the Synthesize--Reflect--Finalize path performs
revision ($K*\phi$): the Coordinator preserves or revises the
preliminary hypothesis using the calibrated evidence.
The final output instantiates the tuple: \texttt{is\_same\_author} $\in
\{\text{true},\text{false}\}$, confidence $c \in [0,1]$, evidence chain
$\mathcal{E}$, verdict reasoning, and dissenting opinions or caveats.

\subsection{Expert Agents}
\label{sec:experts}

Four Expert Agents provide complementary views of authorship evidence and
implement expansion by producing dimension-specific reports for the
Coordinator.
Each agent receives $(x_1, x_2)$ and produces a structured report $e_i$---encoding 
dimension-level similarity, confidence, and confounder risks---that the Coordinator 
assembles into the evidence chain $\mathcal{E} = \{e_1,\ldots,e_k\}$. Each agent 
may also invoke dimension-relevant tools via MCP-based interfaces, external APIs, 
or local functions to extract code features that sharpen its evidence assessment.

\vspace{2pt}
\noindent\textbf{Layout Agent}---analyzes indentation, spacing regularities,
delimiter placement, and comment layout. These features capture formatting
habits and tooling preferences that may remain consistent across code samples.

\noindent\textbf{Lexical Agent}---examines identifier naming conventions,
character-level regularities, and token distributions. These features reflect
preferences in naming style, code expressiveness, and lexical choice.

\noindent\textbf{Syntactic Agent}---analyzes structural organization,
control-flow shapes, and AST-level patterns. In cross-language settings, it
focuses on more transferable structural properties, such as nesting depth and
cyclomatic complexity, rather than language-specific syntax.

\noindent\textbf{Pattern Agent}---captures higher-level programming habits,
including helper-function usage, decomposition strategies, defensive checks,
edge-case handling, and algorithmic preferences. We introduce this agent to
model authorial signals that are less tied to surface syntax and may better
transfer across heterogeneous code pairs.

\textbf{Pattern Agent's Distinctive Role.} The Pattern Agent is especially
important for cross-language verification, i.e., when $x_1$ and $x_2$ are
written in distinct programming languages.
Layout, lexical, and syntactic cues can change substantially when the same
author writes in different programming languages, because each language imposes
different syntax, libraries, and idioms.
In contrast, programming-pattern evidence reflects how a developer decomposes
problems, structures reusable logic, and handles exceptional
cases~\cite{kalgutkar2019code}.
MACAA uses this difference in reliability during contraction: it
down-weights evidence dimensions that are likely distorted by
language-specific factors while preserving more stable pattern-level evidence.
This follows the minimal change principle of belief revision: adjust only
unreliable parts of the belief state rather than replacing the whole authorship
hypothesis.

\section{Experiments}
\label{sec:experiments}

We evaluate MACAA to address four research questions: \textbf{RQ1:} Does belief revision-based multi-agent reasoning outperform existing baselines on same-language verification? \textbf{RQ2:} Can MACAA maintain robust performance on cross-language verification across varying linguistic distance levels? \textbf{RQ3:} What are the contributions of individual belief revision operations? \textbf{RQ4:} How do expert agent weights adapt across same-language and cross-language settings?

\subsection{Experimental Setup}
\label{sec:setup}

\paragraph{Datasets.}
We construct evaluation suites spanning competitive-programming and open-source domains. GCJ-C++~\cite{quiring2019misleading} contains 204 authors and 1,632 C++ files from Google Code Jam. GitHub-C~\cite{kalgutkar2019code} (67 authors, 2,072 files) and GitHub-Java~\cite{kalgutkar2019code} (40 authors, 2,827 files) add real-world diversity. CodeNet~\cite{puri2021codenet} contains 13.9 million submissions in 55 languages. Table~\ref{tab:dataset} summarizes the dataset statistics. Together, these datasets enable comprehensive evaluation across single-language and cross-language settings.

\paragraph{Implementation Details.}
We deploy MACAA using Qwen3-80B-Instruct via vLLM on 2 NVIDIA H200 GPUs. Full hyperparameters are provided in Appendix~\ref{sec:appendix-implementation}.

\paragraph{Baselines.}
We compare against eight methods spanning supervised, zero-shot, and multi-agent paradigms: (1) Stylo-ML \cite{caliskan2015anonymizing}, classic ML with handcrafted features; (2) Forsee \cite{guo2024enhancing}, deep neural network with expert feature knowledge; (3) CLAVE \cite{alvarez2025clave}, transformer-based contrastive learning; (4) LLM Direct \cite{choi2025can}, zero-shot prompting; (5) LLM+Rationale (chain-of-thought) \cite{wei2022chain}; (6) LLM+Tools (single-agent ReAct) \cite{yao2023react}; (7) MAS-Pipeline \cite{li2023camel}, sequential multi-agent without belief revision; (8) MAS-Debate \cite{du2024improving}, parallel multi-agent with debate but no explicit belief operations. Supervised baselines (1)--(3) are trained per their original papers and applied to our evaluation pairs without retraining or fine-tuning; MACAA requires no training by design. Complete training configurations in Appendix~\ref{sec:appendix-protocol}. LLM-based baselines (4)--(8) use a unified backbone (Qwen3-80B-Instruct, temperature 0.1) with identical tool budgets; see Appendix~\ref{sec:appendix-implementation}.

\paragraph{Evaluation Setup.}
We construct balanced evaluation sets through stratified random sampling, measuring performance via F1 score and confusion matrices. Same-language pairs use approximately 1:1 same/different author ratios; cross-language pairs use 200 pairs per language pair. All methods are evaluated on identical test samples.

\begin{table}[t]
\centering
\def\arraystretch{1.15}
\setlength{\tabcolsep}{4pt}
\small
\resizebox{\linewidth}{!}{
\begin{tabular}{lcccc}
\Xhline{1.2pt}
\ccd{Dataset} & \ccd{Language} & \ccd{\# Authors} & \ccd{\# Codes} \\
\Xhline{1.2pt}
\cc{GCJ C++}   & \cc{C++}  & \cc{204}    & \cc{1,632}   \\
GitHub Java   & Java      & 40          & 2,827        \\
\cc{GitHub C} & \cc{C}    & \cc{67}     & \cc{2,072}   \\
CodeNet-10M   & Multi     & 4,048       & 102,350      \\
\Xhline{1.2pt}
\end{tabular}%
}
\caption{Overview of datasets used in experiments.}
\label{tab:dataset}
\end{table}

\subsection{Main Results}
\label{sec:main_results}

\subsubsection{Same-Language Verification (RQ1)}

We evaluate on 500 balanced pairs from each dataset (GCJ C++, GitHub Java, GitHub C). Table~\ref{tab:same-lang-results} presents F1 score and confusion matrices.

Table~\ref{tab:same-lang-results} presents F1 score and confusion matrices. Supervised baselines achieve moderate but stable F1 (63.35\%--70.31\%). LLM Direct shows extreme variance (22.06\%--78.55\%); LLM+Rationale improves slightly (36.30\%--82.79\%), while LLM+Tools narrows the gap (66.50\%--78.53\%). Multi-agent baselines without belief revision perform strongly on GitHub Java (MAS-Pipeline 83.95\%, MAS-Debate 88.10\%) but degrade sharply on the other two datasets (46.11\%--52.38\%), confirming that agent parallelism alone cannot resolve cross-project conflicts.

MACAA achieves the best F1 on GCJ C++ (89.15\%) and GitHub C (74.84\%), and remains competitive on GitHub Java (85.37\%), where MAS-Debate obtains the highest F1 (88.10\%). MACAA's consistent performance (F1 range: 74.84\%--89.15\%) demonstrates that principled belief revision enables stable verification across heterogeneous datasets.

\begin{table}[t]
\centering
\def\arraystretch{1.15}
\setlength{\tabcolsep}{3pt}
\small 
\begin{tabularx}{\linewidth}{l *{5}{>{\centering\arraybackslash}X}}
\Xhline{1.2pt}
\ccd{Method} & \ccd{F1} & \ccd{TP} & \ccd{FP} & \ccd{TN} & \ccd{FN} \\
\Xhline{1.2pt}
\multicolumn{6}{c}{\cellcolor{headblue}\textbf{GCJ C++}} \\
\hhline{======}
\rowcolor{rowgray}
Stylo-ML & 65.49 & 223 & 208 & 42 & 27 \\
Forsee & 69.29 & 238 & 199 & 51 & 12 \\
\rowcolor{rowgray}
CLAVE & 68.93 & 244 & 214 & 36 & 6 \\
LLM Direct & 59.89 & 112 & 12 & 238 & 138 \\
\rowcolor{rowgray}
LLM+Rationale & 68.59 & 142 & 25 & 225 & 105 \\
LLM+Tools & 71.45 & 229 & 162 & 88 & 21 \\
\rowcolor{rowgray}
MAS-Pipeline & 46.11 & 80 & 17 & 233 & 170 \\
MAS-Debate & 52.38 & 99 & 29 & 221 & 151 \\
\rowcolor{headblue}
MACAA & \textbf{89.15} & \textbf{226} & \textbf{31} & \textbf{219} & \textbf{24} \\
\Xhline{1.2pt}
\multicolumn{6}{c}{\cellcolor{headblue}\textbf{GitHub Java}} \\
\hhline{======}
\rowcolor{rowgray}
Stylo-ML & 63.88 & 168 & 113 & 142 & 77 \\
Forsee & 67.73 & 170 & 97 & 168 & 65 \\
\rowcolor{rowgray}
CLAVE & 69.09 & 152 & 53 & 212 & 83 \\
LLM Direct & 78.55 & 152 & 0 & 265 & 83 \\
\rowcolor{rowgray}
LLM+Rationale & 82.79 & 166 & 0 & 265 & 69 \\
LLM+Tools & 78.53 & 172 & 31 & 234 & 63 \\
\rowcolor{rowgray}
MAS-Pipeline & 83.95 & 170 & 0 & 265 & 65 \\
MAS-Debate & \textbf{88.10} & \textbf{185} & \textbf{0} & \textbf{265} & \textbf{50} \\
\rowcolor{headblue}
MACAA & 85.37 & 175 & 0 & 265 & 60 \\
\Xhline{1.2pt}
\multicolumn{6}{c}{\cellcolor{headblue}\textbf{GitHub C}} \\
\hhline{======}
\rowcolor{rowgray}
Stylo-ML & 66.19 & 233 & 221 & 29 & 17 \\
Forsee & 63.35 & 159 & 93 & 157 & 91 \\
\rowcolor{rowgray}
CLAVE & 70.31 & 225 & 165 & 85 & 25 \\
LLM Direct & 22.06 & 31 & 0 & 250 & 219 \\
\rowcolor{rowgray}
LLM+Rationale & 36.30 & 55 & 0 & 250 & 193 \\
LLM+Tools & 66.50 & 135 & 21 & 229 & 115 \\
\rowcolor{rowgray}
MAS-Pipeline & 46.11 & 80 & 17 & 233 & 170 \\
MAS-Debate & 49.86 & 86 & 9 & 241 & 164 \\
\rowcolor{headblue}
MACAA & \textbf{74.84} & \textbf{177} & \textbf{46} & \textbf{204} & \textbf{73} \\
\Xhline{1.2pt}
\end{tabularx}
\caption{Same-language verification results. F1 score (\%); TP/FP/TN/FN: confusion matrix entries.}
\label{tab:same-lang-results}
\end{table}

\subsubsection{Cross-Language Verification (RQ2)}
\label{sec:cross-lang}

We construct cross-language pairs from CodeNet across three divergence levels: Low (C--C++, C--C\#), Medium (Java--Go, Python--Ruby), and High (C++--Python, Java--Haskell), with 200 pairs per language pair. See Appendix~\ref{sec:appendix-lang-dist} for detailed language pair descriptions. We also test a mixed setting with 500 random pairs spanning Python, Java, Go, and C++ without predetermined combinations.

Table~\ref{tab:cross-lang-results} presents F1 score. Feature-based baselines exhibit inconsistent patterns across divergence levels. LLM-based methods show expected degradation as linguistic distance increases: LLM Direct drops from 45.21\% at Low divergence to 23.61\% at High divergence. The multi-agent baselines underscore the importance of principled belief revision: MAS-Pipeline degrades severely (49.15\% to 11.85\%), and MAS-Debate struggles (52.17\% to 22.38\%), confirming that architectural contribution lies not in multi-agent parallelism per se, but in the belief revision operations that calibrate evidence reliability under conflict.

In contrast, MACAA maintains stable performance across all divergence levels (76.19\% at Low, 70.37\% to 70.75\% at Medium, 69.06\% to 76.39\% at High). Under the mixed setting with unpredictable language combinations, MACAA achieves 80.00\% F1, substantially outperforming all baselines (19.20\% to 50.00\%). This confirms that MACAA's belief revision mechanism enables consistent verification accuracy across controlled divergence levels while maintaining autonomous adaptability to arbitrary heterogeneous encounters in open-world scenarios.

\begin{table*}[t]
\centering
\setlength{\aboverulesep}{0pt} 
\setlength{\belowrulesep}{0pt} 
\def\arraystretch{1.35}
\small
\begin{tabularx}{\textwidth}{l *{7}{>{\centering\arraybackslash}X}}
\Xhline{1.2pt}
\rowcolor{headblue} 
 & \multicolumn{6}{c}{\textbf{Linguistic Distance}} & \\
\cmidrule(lr){2-7} 
\rowcolor{headblue} 
 & \multicolumn{2}{c}{\textbf{Low}} & \multicolumn{2}{c}{\textbf{Medium}} & \multicolumn{2}{c}{\textbf{High}} & \\
\cmidrule(lr){2-3} \cmidrule(lr){4-5} \cmidrule(lr){6-7}
\rowcolor{headblue} 
\multirow{-3}{*}{\textbf{Method}} & C--C++ & C--C\# & Java--Go & Py--Ruby & C++--Py & Java--Hask & \multirow{-3}{*}{\textbf{Mix}} \\
\midrule
\rowcolor{rowgray}
Stylo-ML & 61.92 & 51.12 & 54.41 & 64.56 & 61.32 & 77.30 & 19.20 \\
Forsee & 61.32 & 39.34 & 40.27 & 66.91 & 62.36 & 77.91 & 46.55 \\
\rowcolor{rowgray}
CLAVE & 61.81 & 50.00 & 54.07 & 64.71 & 63.08 & 76.68 & 46.39 \\
LLM Direct & 45.21 & 43.67 & 30.76 & 42.10 & 46.55 & 23.61 & 28.28 \\
\rowcolor{rowgray}
LLM+Rationale & 50.42 & 50.54 & 37.89 & 53.12 & 52.89 & 30.66 & 34.57 \\
LLM+Tools & 58.76 & 39.63 & 52.17 & 59.21 & 62.31 & 49.72 & 50.00 \\
\rowcolor{rowgray}
MAS-Pipeline & 49.15 & 44.44 & 21.69 & 17.48 & 41.07 & 11.85 & 26.41 \\
MAS-Debate & 49.15 & 52.17 & 42.55 & 39.32 & 49.57 & 22.38 & 41.03 \\
\rowcolor{headblue}
\textbf{MACAA} & \textbf{76.19} & \textbf{70.37} & \textbf{70.49} & \textbf{70.75} & \textbf{76.39} & \textbf{69.06} & \textbf{80.00} \\
\Xhline{1.2pt}
\end{tabularx}
\caption{Cross-language verification F1 score (\%) by linguistic distance. 200 pairs per language pair, 500 pairs for Mix.}
\label{tab:cross-lang-results}
\end{table*}

\subsection{Belief Revision Ablation and Expert Contribution Analysis (RQ3--RQ4)}
\label{sec:ablation}

\begin{table}[t]
\centering
\def\arraystretch{1.12}
\setlength{\tabcolsep}{4pt}
\small
\begin{tabular}{>{\centering\arraybackslash}m{2.2cm}ccc}
\Xhline{1.2pt}
\ccd{Configuration} & \ccd{F1 (Same)} & \ccd{F1 (Cross)} & \ccd{EC (\%)} \\
\Xhline{1.2pt}
\cc{Full MACAA} & \cc{\textbf{83.08}} & \cc{\textbf{75.37}} & \cc{66.66~/~24.00} \\
w/o Expansion & 19.35 & 11.11 & N/A~/~N/A \\
\cc{w/o Contraction} & \cc{76.47} & \cc{71.19} & \cc{68.00~/~24.00} \\
w/o Revision & 77.61 & 61.82 & 64.00~/~24.00 \\
\Xhline{1.2pt}
\end{tabular}%
\caption{Belief revision ablation. F1 score (\%); EC: Expert Consensus rate (Same/Cross).}
\label{tab:ablation-compact-v3}
\end{table}

\begin{table}[t]
\centering
\def\arraystretch{1.25}
\setlength{\tabcolsep}{10pt}
\small
\begin{tabular}{lcc}
\Xhline{1.2pt}
\ccd{Agent} & \ccd{Same-Lang} $\bar{w}$ & \ccd{Cross-Lang} $\bar{w}$ \\
\Xhline{1.2pt}
\cc{Layout}   & \cc{0.1709(--0.0591)} & \cc{0.1950(--0.0250)} \\
Lexical   & 0.2869(+0.0096) & 0.2570(--0.0430) \\
\cc{Syntactic} & \cc{0.3115(+0.0415)} & \cc{0.0671(--0.0129)} \\
Pattern   & 0.2279(+0.0079) & 0.4810(+0.0810) \\
\Xhline{1.2pt}
\end{tabular}
\caption{Adaptive agent weights across verification settings. Parentheses denote changes from initial weights.}
\label{tab:expert-analysis}
\end{table}

We conduct RQ3 and RQ4 under the same evaluation settings: same-language verification with 500 randomly sampled GCJ C++ code pairs, and cross-language verification with 200 randomly sampled C--C++ code pairs. RQ3 ablates the belief-revision operations to measure their contribution, while RQ4 analyzes adaptive agent weights to examine how MACAA calibrates evidence reliability across verification settings.

\textbf{Belief Revision Ablation (RQ3).}
Table~\ref{tab:ablation-compact-v3} reports F1 score and Expert Consensus (EC), where EC is the proportion of samples on which all four Expert Agents make the same vote. Full MACAA achieves 83.08\% F1 on same-language pairs and 75.37\% F1 on cross-language pairs. Removing Expansion causes severe degradation, reducing F1 to 19.35\% and 11.11\%, which confirms that single-pass reasoning cannot capture multi-dimensional authorship evidence; note that Contraction and Revision both depend on evidence produced by Expansion, so removing Expansion disables the entire revision chain. Removing Contraction lowers same-language F1 to 76.47\%, while cross-language F1 drops slightly to 71.19\%, showing that contraction contributes to cross-language calibration by down-weighting unreliable evidence. Removing Revision reduces F1 to 77.61\% and 61.82\%, showing that the final hypothesis update is especially important when expert evidence is conflicting. The much lower cross-language EC, 24.00\%, further indicates stronger disagreement among evidence dimensions in heterogeneous verification.

\textbf{Expert Agent Contribution (RQ4).}
Table~\ref{tab:expert-analysis} reports the adaptive weight distribution across the same verification settings. In same-language verification, the Syntactic and Lexical agents receive the largest weights, 0.3115 (+0.0415) and 0.2869 (+0.0096), respectively. This suggests that when two samples share the same programming language, token choices and structural forms provide reliable authorship cues. The Layout weight decreases to 0.1709 (-0.0591), indicating that formatting signals are less stable and may be affected by editors or project conventions. The Pattern Agent increases slightly to 0.2279 (+0.0079), but it is not dominant in this setting. In cross-language verification, the Pattern Agent becomes the most reliable dimension, rising to 0.4810 (+0.0810), because decomposition strategies, helper-function usage, and edge-case handling are less tied to surface syntax. By contrast, Syntactic weight drops to 0.0671 (-0.0129), while Lexical and Layout weights decrease to 0.2570 (-0.0430) and 0.1950 (-0.0250), reflecting the instability of language-specific syntax, tokens, and formatting. These results validate MACAA's contraction mechanism: the framework calibrates dimension-level reliability by weakening unreliable or language-induced evidence while preserving useful signals for later revision.

\paragraph{Interpretability and Auditable Reasoning.}
MACAA produces complete belief revision traces with explicit evidence provenance, enabling human-in-the-loop audit. A representative case is provided in Appendix~\ref{sec:appendix-interpretability}.

\section{Conclusion}
\label{sec:conclusion}

We presented MACAA, an AGM-inspired multi-agent framework for training-free code authorship verification. By framing verification as iterative belief update rather than static classification, MACAA addresses the fundamental limitation of supervised approaches that require fixed author sets and fail in open-world deployment. Our architecture implements expansion, contraction, and revision through a Coordinator Agent and four Expert Agents, enabling robust verification across heterogeneous code pairs with complete evidence provenance. Experiments demonstrate competitive or superior performance across same-language and cross-language tasks, with strong cross-dataset stability. MACAA establishes belief revision as a practical foundation for interpretable, forensically auditable authorship verification.

\section{Limitations}
\label{sec:limitations}

We identify three limitations. First, all experiments employ Qwen3-80B-Instruct. We adopt a single backbone to isolate architectural effects from model-capability effects: when the same model achieves 22.06\% F1 under direct prompting yet 74.84\% F1 under MACAA on identical GitHub C pairs, the improvement is attributable to the belief-revision architecture rather than to the underlying LLM. Nevertheless, whether these gains generalize to other LLM families and parameter scales remains to be validated. Second, the cross-language evaluation covers only high-resource languages (Python, Java, Go, C, C++, Ruby, Haskell), and same-language datasets come from competitive programming and open-source domains; performance on low-resource languages, obfuscated code, or enterprise codebases is unknown. Third, iterative multi-agent belief revision requires more LLM calls than direct prompting; we leave systematic quantification of this cost--accuracy trade-off to future work.

\bibliography{custom}

\appendix

\section{Experimental Setup and Baseline Training}
\label{sec:appendix-protocol}

Following standard practice \cite{kalgutkar2019code}, we construct balanced evaluation sets through stratified random sampling. For same-language evaluation, we sample 500 pairs per dataset with approximately 1:1 same/different author ratios. For cross-language evaluation, we sample 200 pairs per language pair in the stratified setting and 500 pairs in the mixed setting. Under approximately 1:1 class balance, F1 provides unbiased assessment of discrimination capability. All code samples are preprocessed to remove explicit author identifiers while preserving stylistic features.

We adopt F1 score as the primary metric: $\text{F1} = 2 \times \frac{\text{Precision} \times \text{Recall}}{\text{Precision} + \text{Recall}}$. The confusion matrix definitions are: TP (True Positive): same-author pairs correctly predicted; TN (True Negative): different-author pairs correctly predicted; FP (False Positive): different-author pairs incorrectly predicted as same-author; FN (False Negative): same-author pairs incorrectly predicted as different-author. 

We train supervised baselines (Stylo-ML, Forsee, CLAVE) following the original paper configurations while ensuring fair comparison with MACAA. For Stylo-ML and Forsee, we maintain the original training data scale but use code samples from our evaluation domains to ensure these methods learn task-relevant distribution. For CLAVE, we conduct pretraining on a corpus matching the original paper size but containing all language types used in our experiments, followed by feature extraction and classification. All baselines are trained with their original hyperparameters and validation protocols reported in respective papers.

This training strategy serves two purposes. First, it ensures fair comparison by allowing supervised methods to learn from the same task distribution as MACAA, eliminating distribution mismatch as a confounding factor. Second, and more critically, it enables demonstration of open-world generalization challenges: supervised models trained on specific author sets and language combinations encounter unseen authors and cross-language pairs during testing, revealing performance degradation inherent to closed-world assumptions. Strictly speaking, a fully fair comparison with MACAA would require supervised baselines to operate without any training; however, we retain their original supervised training setup to explicitly showcase the performance gap between training-dependent methods and training-free approaches when both face open-world test scenarios with out-of-distribution samples.

\section{Cross-Language Language Pair Descriptions}
\label{sec:appendix-lang-dist}

\noindent\textbf{Low distance} (C--C++, C--C\#): The low-distance group includes C--C++ and C--C\#. These pairs are close enough that many surface and structural cues remain comparable. C and C++ share a direct C-family lineage, similar expression syntax, brace-based block structure, and common imperative control-flow forms. C and C\# are not identical in type system or runtime model, but they still share C-style lexical conventions, block delimiters, operator syntax, and common loop and branching forms. These pairs therefore test whether verification can use residual structural similarity without confusing language-family similarity with authorship similarity.

\noindent\textbf{Medium distance} (Java--Go, Python--Ruby): The medium-distance group includes Java--Go and Python--Ruby. These pairs share some broad programming conventions but differ in the details that often shape code style. Java and Go are both statically typed, brace-based, and commonly used for modular program organization, yet they differ in object orientation, interface use, error handling, and standard idioms. Python and Ruby are both dynamically typed scripting languages with compact syntax and high-level library use, but they differ in block notation, method-call style, iterator idioms, and object conventions. These pairs test verification under partial transfer: some abstract habits, such as decomposition and naming preferences, may remain visible, while many syntactic and idiomatic cues no longer align directly.

\noindent\textbf{High distance} (C++--Python, Java--Haskell): The high-distance group includes C++--Python and Java--Haskell. These pairs place the verifier under stronger language-induced shift. C++ and Python differ in typing discipline, memory and resource management, syntax density, standard-library style, and typical abstraction mechanisms. Java and Haskell differ even more sharply in programming paradigm, with Java centered on imperative object-oriented structure and Haskell centered on functional composition, algebraic data types, and expression-oriented design. In these settings, surface markers such as keywords, delimiters, and local syntax provide weak cross-language evidence. The main question is whether MACAA can down-weight such unstable cues and rely more on transferable author habits, including problem decomposition, helper-function design, edge-case handling, and recurring semantic preferences.

\section{Interpretability and Auditable Reasoning}
\label{sec:appendix-interpretability}

This appendix presents the complete reasoning trace of MACAA on a cross-language case (Python vs C\texttt{++}, ground truth: same author). Each box shows actual system output at that stage. The purpose of this trace is to demonstrate how MACAA exposes every intermediate decision to human inspection: the auditor can verify each expert report against the source code, check whether confidence adjustments are justified, and identify points where the automated reasoning may require human override. The trace is not presented as a validated forensic conclusion, but as an illustration of the auditable interface that MACAA provides for human-in-the-loop verification.

\paragraph{Configuration \& Preliminary Review.}
Cross-language detected (Python vs C\texttt{++}); dimension weights adjusted and incompatible tools disabled. The Coordinator then establishes a Bayesian prior---not a vote.

\begin{lstlisting}[style=promptblock,
  caption={Coordinator: configuration and preliminary review.}]
[INIT] cross-language: CODE1=Python, CODE2=C++
  weights: Layout=0.22, Lexical=0.30,
    Syntactic=0.08, Pattern=0.40
  disabled: token_freq/ngram/abstract,
    ast_node/path/construct/dolos, api_idiom

[PRELIMINARY] tendency: same_author
  confidence: 0.62
  reasoning:
    1. naming: both codes use short,
       lowercase-dominant identifiers
       (avg_len<3), suggesting consistent
       personal naming compression habit.
    2. structure: both adopt flat,
       single-block scripts without
       helper functions or abstractions.
    3. comment: both are comment-free,
       aligning with rapid-competition
       authoring style.
    4. confounders: competitive template,
       language_syntax may mimic
       author-level consistency.
\end{lstlisting}

\paragraph{Expert Evidence.}
Four Expert Agents analyze complementary dimensions in parallel via ReAct tool loops.

\begin{lstlisting}[style=promptblock,
  caption={Layout and Lexical Agent evidence.}]
[LAYOUT] s=0.32, conf=0.58 -> different
  Python: space (37/37), avg_indent=7.28,
    indent_std=3.37, comma_tight=45,
    comma_space=27, no comments.
    indent_switch=0.59 (moderate),
    line_len_std=15.55, K&R absent.
  C++: tab (85/85), avg_indent=1.37,
    indent_std=0.76, K&R braces
    (same_line_opener=11), no comments.
    indent_switch=0.24 (consistent),
    line_len_std=7.97, comma_space=3.
  Verdict: Tab vs Space 100% opposite;
    avg_indent 5.3x divergence;
    C++ format 2x more stable.

[LEXICAL] s=0.52, conf=0.67 -> uncertain
  Python: lower=96.7%, snake=3.3%,
    avg_len=2.82, unique_ratio=31%.
    Top: flag(7), s(5), key(5),
    h(4), g(4), p(4), data(4).
    84% identifiers <= 4 chars.
    kw: if(7), break(4), while(1).
  C++: lower=93.9%, snake=6.1%,
    avg_len=2.08, unique_ratio=17%.
    Top: c(25), d(24), i(9), j(9).
    No custom funcs/classes.
  Naming style consistent (lower>93%,
    avg_len<3); specific names
    language-driven (high confounder).
\end{lstlisting}

\begin{lstlisting}[style=promptblock,
  caption={Syntactic and Pattern Agent evidence.}]
[SYNTACTIC] s=0.45, conf=0.49 -> uncertain
  Python: depth=16, branching_std=2.34,
    656 nodes, avg_branching=1.02.
  C++: depth=12, branching_std=2.07,
    1130 nodes, avg_branching=1.01.
  Flat trees structurally similar;
    AST not cross-language alignable.
  Confounder: competition template
    forces flat structure in both.

[PATTERN] s=0.42, conf=0.48 -> uncertain
  Python: 0 funcs, 7 if/2 for loops,
    ccn=7, short_temp=0.00.
    Zero assertions, zero constants,
    zero helpers, zero classes.
  C++: 1 main (ccn=14), 5 if/5 loops,
    short_temp=0.19, zero helpers.
    Zero assertions, zero constants.
  Both: no assert/const/helper pattern;
    func count language-driven (0 vs 1).
\end{lstlisting}

\paragraph{Synthesis.}
The Coordinator synthesizes all expert reports and detects a high-severity conflict.

\begin{lstlisting}[style=promptblock,
  caption={Synthesis Round 1 output.}]
tendency: different_author (conf=0.55)
  layout: different (0.58)
    Tab/Space 100% opposite.
    avg_indent 7.28 vs 1.37 (5.3x).
    K&R vs non-K&R conflict.
    C++ format 2x more stable.
  lexical: uncertain (0.67)
    naming style consistent
    (lower>93%, avg_len<3).
    Specific names language-driven.
    High confounder risk.
  syntactic: uncertain (0.65)
    avg_branching=1.02 vs 1.01.
    Flat trees similar but AST not
    cross-language alignable.
    Confounder: competition template.
  pattern: uncertain (0.61)
    zero helpers/assert/const both.
    Func count (0 vs 1) driven
    by language ecosystem.

critical_conflict: layout vs lexical
  (Tab/Space opposition vs
   naming consistency)
action: START_DEBATE
  lexical (proponent) vs layout
\end{lstlisting}

\paragraph{Structured Debate.}
The Layout--Lexical conflict triggers a 3-turn debate. Lexical argues the \texttt{flag}$\to$\texttt{is\_half} semantic chain is author-stable; Layout counters that Tab/Space is a low-level fingerprint with $>$85\% persistence.

\begin{lstlisting}[style=promptblock,
  caption={Debate: Lexical vs Layout (3 turns).}]
TURN 1 [Pro-Lexical]:
  flag->is_half/is_half2 semantic chain
  maps identical boolean-levelization
  logic. Confirmed in source; not
  template-replaceable.
  Naming style consistent: lower>93%,
  avg_len<3 across both languages.
  Low unique ratio (31% vs 17%)
  reflects same compression habit.

TURN 2 [Con-Layout]:
  Tab/Space 100% opposite is an
  author-stable fingerprint.
  Literature: >85% persistence rate.
  avg_indent 5.3x divergence is
  not explainable by language alone.
  K&R vs non-K&R also conflicts.

TURN 3 [Rebuttal]:
  indent adapts to language ecosystem.
  Python forces space; C++ CP defaults
  to Tab. No evidence of systematic
  author-level switching.
  flag->is_half source-code verified:
  flag at time-judge entry, is_half
  at same logic position, is_half2
  is flag's override logic.

Result: Lexical promoted to
  same_author (conf=0.68).
  Syntactic conf lowered: 0.65->0.49.
\end{lstlisting}

\paragraph{Recheck and Final Decision.}
Layout rechecked twice; $s{=}0.32$ confirmed both times with no counter-evidence. The Coordinator finalizes.

\begin{lstlisting}[style=promptblock,
  caption={Layout recheck and final decision.}]
[RECHECK 1] layout: s=0.32 (same)
  Tab/Space gap confirmed.
  indent_switch=0.59 vs 0.24.
  C++ format 2x more stable.
  No new counter-evidence found.
[RECHECK 2] layout: s=0.32 (same)
  Gap persists; no counter-evidence.
  Confounder strengthened:
  language ecosystem fully accounts
  for Tab/Space opposition.
  Python forces space; C++ CP
  defaults to Tab.

[FINALIZE] same_author (conf=0.79)
  weights: Lex=0.35, Lay=0.25,
    Syn=0.20, Pat=0.20
  rationale:
  1. Lexical (s=0.52, conf=0.68):
     flag->is_half chain is stable,
     ecosystem-independent.
     Source-code verified: flag at
     time-judge entry, is_half at
     same logic position.
  2. Layout (s=0.32) downweighted:
     Tab/Space attributed to language
     ecosystem (Py=space, C++=Tab).
     Rechecked 2x, no counter found.
  3. Syntactic/Pattern uncertain but
     do not contradict.
  opponent notes:
     Tab/Space persistence >85% in
     literature; but no evidence
     author systematically switched.
  process: 4 rounds, 1 debate,
    2 rechecks, 35/40 LLM calls.
  anchored: PRELIMINARY same@0.62.
\end{lstlisting}

The final decision (\texttt{same\_author}, 0.79) agrees with the ground truth. The trace above demonstrates how MACAA resolves cross-dimensional conflicts through transparent belief revision: Layout's Tab/Space opposition, the strongest counter-signal, is ultimately attributed to language ecosystem conventions, while Lexical's naming patterns (notably the \texttt{flag}$\to$\texttt{is\_half} mapping confirmed in source code) persist through debate and recheck as author-stable evidence. Each intermediate decision, including confidence adjustments and tool outputs, is logged and auditable.

\section{Implementation Details}
\label{sec:appendix-implementation}

We deploy MACAA using Qwen3-80B-Instruct via vLLM (v0.6.6) with tensor parallelism across 2 NVIDIA H200 GPUs, at temperature 0.1 and max\_tokens 4096. The Coordinator runs up to 4 rounds with at most 2 debate sessions (3 turns each), capped at 40 LLM calls per sample. Initial dimension weights differ by setting: for same-language verification, Layout, Lexical, Syntactic, and Pattern receive 0.23, 0.28, 0.27, and 0.22; for cross-language, they receive 0.22, 0.30, 0.08, and 0.40. Pairs with score $\geq 0.62$ are classified as same-author, those $\leq 0.38$ as different-author, and scores in $(0.38, 0.62)$ as uncertain. Scores $> 0.70$ allow early termination; scores $\leq 0.50$ trigger a targeted recheck.

All LLM-based baselines (LLM Direct, LLM+Rationale, LLM+Tools, MAS-Pipeline, MAS-Debate) are re-implemented under the same backbone (Qwen3-80B-Instruct, temperature 0.1, max\_tokens 4096) and evaluated on identical test pairs. LLM Direct uses a single zero-shot prompt; LLM+Rationale adds chain-of-thought; LLM+Tools enables a single-agent ReAct loop with the same tool set as the Expert Agents. MAS-Pipeline and MAS-Debate replicate the multi-agent architectures from their respective original papers but substitute the backbone model with Qwen3-80B-Instruct and allow up to 40 LLM calls per sample to match MACAA's budget. This controlled setup ensures that performance differences reflect architectural choices rather than backbone model capabilities.

\section{Agent Prompts}
\label{sec:appendix-agent-prompts}

This appendix presents all prompt specifications for MACAA's multi-agent system. We organize prompts by agent type: \S\ref{sec:f-layout}--\ref{sec:f-pattern} cover the four Expert Agents (layout, lexical, syntactic, pattern), each with a ReAct phase for feature extraction and a Comparison phase for pairwise evidence assessment; \S\ref{sec:f-coordinator} presents the Coordinator Agent across its key operational stages (preliminary review, synthesis, strategic decision, reflection, truth-seeking debate, and final decision). All prompts require JSON-only output for programmatic integration; \texttt{confidence} and \texttt{similarity\_score} are normalized to $[0,1]$.

\subsection{Layout Agent}
\label{sec:f-layout}

Targets whitespace, delimiter placement, comment positioning, and formatting stability. ReAct loop with 4 tools, \texttt{max\_steps}$\geq$4.

\subsubsection{ReAct Phase}

\begin{lstlisting}[style=promptblock,caption={Layout Agent ReAct system prompt.}]
You are a "Layout Expert Agent." Your task: perform interpretable,
reproducible, comparable layout-style analysis on a code sample.
Obtain observations via callable tools, organize into evidence
for the Coordinator's authorship attribution.

Goal:
- Autonomously decide next tool (or stop) with causal rationale.
- Each step maximizes information gain, reduces uncertainty.
- Output is structured, interpretable, traceable to observations.

Tools (provided by program):
- extract_whitespace_profile
- extract_delimiter_layout_profile
- extract_comment_layout_profile
- extract_format_stability_profile
One tool per step, or stop (stop=true).

Coverage Constraint:
- max_steps >= 4: all 4 tools must be invoked at least once.
- No repeated calls to the same tool.
- max_steps < 4: prioritize most informative/complementary tools.

ReAct Methodology: Hypothesis-Verify-Update
Each tool selection must answer:
1) Current uncertainty? (which layout dims lack/conflict evidence)
2) Candidate tool's new info? (expected distinguishing signals)
3) Why now? (max info gain, complementary, avoid repetition)

ReAct Structure per step:
1) Thought: current uncertainty dimension, expected signals,
   causal link from previous observation.
2) Action: select tool. Priority: uncovered > complementary >
   conflict resolution.
3) Observation: convert output to 1-3 signals. Assess
   template/task influence; downweight if affected.
4) Stop: when coverage met or budget exhausted. Output evidence:
   summary (one-line style portrait), signals (per dimension),
   confidence (0-1, stability confidence, not same-author).

Output (strict JSON only, no text/markdown/fences):
Continue: {"thought":"...",
  "action":{"type":"tool",
    "name":"tool_name"},
  "stop":false}
Stop: {"thought":"...","action":
  {"type":"stop"},
  "stop":true,
  "evidence":{
    "summary":"...",
    "signals":["..."],
    "confidence":0.0}}
\end{lstlisting}

\subsubsection{Comparison Phase}

\begin{lstlisting}[style=promptblock,caption={Layout Comparator system prompt.}]
You are a LayoutComparator in a code authorship attribution system.
Your task: judge whether two layout profiles come from the same author.

Features to compare:
1. whitespace_profile: avg_indent, tab/space lines, avg_line_length,
   empty_line_ratio, trailing_space_lines, indent_std
2. delimiter_layout_profile: control_space_before_paren,
   control_tight_before_paren, comma_space/tight,
   same_line_block_opener, next_line_block_opener
3. comment_layout_profile: comment_line_ratio, inline_comments,
   standalone_comments, doc_comments
4. format_stability_profile: indent_switch_rate, line_length_std

Key judgment principles:
- Indentation and spacing preferences are strong author signals.
- Delimiter formatting habits (if(x) vs if (x)) are stable.
- Comment style aids judgment but content is task-influenced.
- Large code-size differences distort absolute metrics; focus on ratios.
- Layout is HIGH-confounder in competitive programming:
  High-risk: 2/4-space indent, K&R braces, compact formatting,
    sparse comments, standard comma spacing, common CP patterns.
  Low-risk: uncommon internal formatting combos, rare indentation
    combos, unique empty-line rhythm, stable trailing spaces.

Output: {similarity_score, confidence, matches[],
  differences[], author_stable_matches[],
  template_or_task_matches[], high_risk_confounders[],
  signal_quality, confounder_risk, reasoning}
\end{lstlisting}

\subsection{Lexical Agent}
\label{sec:f-lexical}

Examines token distributions, naming conventions, identifier morphology, and abstract lexical templates. ReAct loop with 5 core tools, \texttt{max\_steps}$\geq$5.

\subsubsection{ReAct Phase}

\begin{lstlisting}[style=promptblock,caption={Lexical Agent ReAct system prompt.}]
You are a "Lexical Expert Agent." Perform interpretable, reproducible,
comparable lexical-style analysis. Tools:
- extract_token_frequency_profile
- extract_token_ngram_profile
- extract_char_ngram_profile
- extract_identifier_style_profile
- extract_abstract_lexical_profile
Coverage: max_steps >= 5, all 5 tools invoked at least once.
ReAct methodology, step structure, output format: identical to
Layout Agent ReAct phase (Section E.1).
\end{lstlisting}

\subsubsection{Comparison Phase}

\begin{lstlisting}[style=promptblock,caption={Lexical Comparator system prompt.}]
You are a LexicalComparator. Judge whether two lexical profiles come from the same author.

Features to compare:
1. token_frequency_profile: keyword_ratio, identifier_ratio,
   operator_ratio, punctuation_ratio, token_top
2. token_ngram_profile: token_bigrams, abstract_token_trigrams,
   longest_repeated_sequence
3. char_ngram_profile: char_4gram, char_5gram
4. identifier_style_profile: identifier_cases, avg_length,
   unique_ratio, digit_ratio, underscore_ratio
5. abstract_lexical_profile: abstract distributions + bigrams

Key principles:
- Naming style = strong author signal (snake_case vs camelCase,
  identifier length, abbreviation habits). Stable across projects.
- Abstract templates > concrete tokens. "if(ID)" vs "if(ID==NUM)".
- Same-author/different-problem: trust identifier_style,
  abstract_lexical, char_ngram over raw token_top.
- Competitive homogenization: keyword ratios converge in CP code;
  without naming-level overlap, similarity <= ~0.62.

High-risk: raw char-ngrams from boilerplate, top tokens (include/
  int/for/cin/cout/return), contest templates, short-loop i/j/k.
Low-risk: identifier morphology + casing, abstract lexical templates,
  recurring naming discipline, underscore_ratio + avg_length.
\end{lstlisting}

\subsection{Syntactic Agent}
\label{sec:f-syntactic}

Examines AST structure, tree shape, and construct usage. ReAct loop with 4 tools; Coordinator additionally invokes Dolos (Tree-sitter + k-gram fingerprinting) outside the loop.

\subsubsection{ReAct Phase}

\begin{lstlisting}[style=promptblock,caption={Syntactic Agent ReAct system prompt.}]
You are a "Syntactic Expert Agent." Perform interpretable, reproducible,
comparable syntactic/structural style analysis. Tools:
- extract_ast_node_profile (degraded mode: keyword pseudo-nodes)
- extract_ast_path_profile (degraded mode: abstract token bigrams)
- extract_tree_shape_profile (degraded mode: bracket-based depth)
- extract_construct_usage_profile (if/for/while/switch/return/...)
Coverage: max_steps >= 4, all 4 tools invoked at least once.
External tool (NOT in ReAct loop):
- analyze_with_dolos: Tree-sitter + k-gram fingerprinting with
  variable masking. Returns dolos_similarity, total_overlap,
  longest_fragment. Treat as supplementary reference.
ReAct methodology, step structure, output format: identical to
Layout Agent ReAct phase (Section E.1).
\end{lstlisting}

\subsubsection{Comparison Phase}

\begin{lstlisting}[style=promptblock,caption={Syntactic Comparator system prompt.}]
You are a SyntacticComparator. Judge whether two syntactic profiles
come from the same author.

Features to compare:
1. ast_node_profile: node type ratios (degraded mode possible)
2. ast_path_profile: parent_child_pairs, sibling_pairs
3. tree_shape_profile: max_depth, avg_branching, branching_std,
   node_count
4. construct_usage_profile: if/for/while/switch/return ratios
5. [Optional] Dolos: dolos_similarity, total_overlap,
   longest_fragment (reference only)

Key principles:
- AST paths + context = core author signals.
- Tree shape = structural thinking (nested vs flat).
- Control-structure prefs (for vs while, early return) = stable.
- Size differences: compare RATIOS, not absolutes.
- Degraded mode: reduce confidence.
- Similarity should not drop below ~0.42 for size mismatches alone.

High-risk: standard DFS/BFS/Dijkstra/DP boilerplate, problem-driven
  control-flow shifts, for-loop dominance, node_count.
Low-risk: recurring AST organization beyond templates, stable
  parent_child/sibling pair prefs, branching_std (normalized).
\end{lstlisting}

\subsection{Pattern Agent}
\label{sec:f-pattern}

Captures function decomposition, control strategy, API/idiom usage, and semantic naming patterns. ReAct loop with 4 tools, \texttt{max\_steps}$\geq$4.

\subsubsection{ReAct Phase}

\begin{lstlisting}[style=promptblock,caption={Pattern Agent ReAct system prompt.}]
You are a Pattern Expert Agent. Extract "programming pattern" evidence
from a single code sample. Tools:
- extract_lizard_function_profile: function_count, avg/max NLOC,
  avg/max cyclomatic complexity, avg parameter count,
  complexity_concentration_ratio, main_like_ccn_ratio
- extract_control_strategy_profile: guard_clause_count,
  guard_if_ratio, recursive_function_hints, loop_count, if_count
- extract_api_idiom_profile: api_families (collections/sorting/
  io/assertions/exceptions), plugin_flags (competitive_header,
  fast_io_optimization)
- extract_semantic_habit_profile: short_temp_ratio,
  helper_name_ratio, uppercase_constant_ratio, assert_like_count
Coverage: max_steps >= 4, all 4 tools invoked at least once.
Recommended order: lizard -> control -> api -> semantic.
lizard_function_profile is the primary anchor for Pattern dimension.
If different tools give conflicting signals, state the conflict
explicitly rather than forcing consistency.
ReAct methodology, step structure, output format: identical to
Layout Agent ReAct phase (Section E.1).
\end{lstlisting}

\subsubsection{Comparison Phase}

\begin{lstlisting}[style=promptblock,caption={Pattern Comparator system prompt.}]
You are a PatternComparator. Judge whether two programming-pattern
profiles come from the same author.

Features to compare:
1. function_metric_profile: function_count, avg_lines_per_function,
   return_per_function, avg_line_length
2. control_strategy_profile: guard_if_ratio, recursive_function_hints,
   loop_count, if_count
3. api_idiom_profile: api_families; plugin_flags
4. semantic_habit_profile: short_temp_ratio, helper_name_ratio,
   uppercase_constant_ratio, assert_like_count

Key principles:
- Function size + organization = stable.
- Control strategy = core author signal (guard clause, recursion).
- Semantic habits = strong signals: temp variable naming (i/j/k vs
  x/y/z), helper naming, constant style.
- Code-size differences: compare RATIOS not absolutes.
- Same-author/different-problem: similarity stays 0.45-0.68 when
  only scale differs; never <0.35 for "one simple, one complex."

High-risk: contest headers, fastio, container combos, common temp
  vars (i/j/k/tmp/ans), function_count/loop_count (problem-driven).
Low-risk: function decomposition habits, complexity distribution,
  helper_name_ratio, uppercase_constant_ratio.
\end{lstlisting}

\subsection{Coordinator Agent}
\label{sec:f-coordinator}

Orchestrates a 10-state machine. Six states are LLM-driven.

\subsubsection{Preliminary Review}

\begin{lstlisting}[style=promptblock,caption={Preliminary Review prompt (same-language).}]
You are the coordinator. Holistic first-pass review; NOT the final
verdict. Contest Code Context: GCJ/KickStart/ACM-ICPC/AtCoder.
IGNORE contest artifacts:
  C/C++: bits/stdc++.h, using namespace std, ios::sync, long long,
    STL containers, macros.
  Python: import sys/math, sys.setrecursionlimit,
    input=sys.stdin.readline, list comps.
  Java: import java.util.*, public class Main, FastScanner.
  Go: package main, fmt.Println, bufio.NewReader.
  Agnostic: short loop vars (i/j/k), main()/solve() scaffolding.

Signal Families (author-stable first):
1. Coding Style  2. Naming  3. Code Structure
4. Control-Flow  5. Comments  6. Language Features
7. Error-Handling  8. Lexical Fingerprints
9. Statistical Cues  10. Idiosyncratic.

Hard Rules: no default to different_author from artifacts;
  mark confounders; balanced same/different.

Output: {overall_first_impression, candidate_style_axes[],
  suspected_confounders[], dimension_routing{layout/lexical/
  syntactic/pattern{priority,why,focus_question}},
  global_questions[], do_not_overtrust[]}
\end{lstlisting}

\begin{lstlisting}[style=promptblock,caption={Preliminary Review prompt (cross-language additions).}]
CRITICAL: CROSS-LANGUAGE mode.
IGNORE: language-specific syntax (keywords, control structures,
  types, stdlib) -- NOT author evidence.
FOCUS: naming conventions, indentation style, spacing habits,
  comment style, code organization.
DO NOT COMPARE: syntax structures, keyword frequencies, AST.

Signal Reliability: Indentation=HIGH, Naming=HIGH, Spacing=HIGH,
  Comment=MED-HI, Organization=MEDIUM, Control-flow=MEDIUM,
  Syntax=LOW, Library=LOW.
Dimension Routing: Layout=MEDIUM, Lexical=HIGH, Syntactic=LOW,
  Pattern=HIGH.
Confidence range: 0.35-0.70.

Output adds: cross_language_mode, detected_languages,
  cross_language_stable_signals.
\end{lstlisting}

\subsubsection{Synthesis}

\begin{lstlisting}[style=promptblock,caption={Synthesis prompt.}]
Coordinator as "Research Manager." Synthesize ExpertReports.
Must answer: (1) Per-dimension tendency + confidence?
(2) Cross-dimension conflicts? (3) Insufficient/degraded dims?
(4) Overall tendency + confidence? (5) What changed?
(6) What remains unanswered?

Core: preliminary_review is PRIOR not VOTE (confirm/weaken/overturn).
"uncertain" preserved, never compressed.
Size differences => prefer "same author, different problem" unless
  strong abstract-layer counter-evidence.
Evidence Chain: high-confounder dims not "strong";
  template_or_task-dominated = weak/moderate;
  only low-confounder author-stable = strong support.

Output: {
  per_dimension_summary[4]{
    dimension, tendency,
    confidence, key_signal,
    is_degraded,
    signal_quality,
    confounder_risk},
  cross_dimension_analysis{
    consensus, conflicts,
    uncertain, max_severity},
  preliminary_update{
    status, why},
  evidence_balance{
    same_author_support,
    different_author_support,
    main_drivers,
    high_risk_confounders},
  overall_assessment{
    tendency, confidence,
    reasoning},
  what_changed,
  remaining_questions}
\end{lstlisting}

\subsubsection{Strategic Decision}

\begin{lstlisting}[style=promptblock,caption={Decision prompt.}]
Strategy Planner selects exactly one action:
1. FINALIZE: evidence sufficient or budget exhausted.
2. RECHECK_DIMENSION: low-confidence/high-impact dimension.
3. START_DEBATE: two dimensions in conflict.
4. ADJUST_WEIGHTS: post-debate/recheck credibility shift.

Priority: no conflict+full evidence > FINALIZE;
  preliminary conflict > RECHECK; two expert dims conflict > DEBATE.
Mandatory: dim-divergence check (dim<0.40 + dim>=0.60 => DEBATE/
  RECHECK). LLM comparison failure => must RECHECK.
Debate participants = real dimensions only
  (layout|lexical|syntactic|pattern).

Output: {action_type, reasoning, params{...}}
\end{lstlisting}

\subsubsection{Reflection}

\begin{lstlisting}[style=promptblock,caption={Reflection prompt.}]
Meta-cognitive reviewer. NOT judging authorship -- assessing whether
current evidence is sufficient for a reliable judgment.

Three mandatory questions:
1. Evidence Sufficiency: all dimensions covered? concrete signals?
2. Conflict Resolution: cross-dimension conflicts explained?
3. Marginal Gain: how much new info could further investigation yield?

Low-Similarity Veto Rule:
- A dimension with similarity_score < 0.40 is "suspect."
- Suspect dims prevent evidence_sufficient unless:
  (a) after debate/recheck, similarity rises to >= 0.45, OR
  (b) the gap is confirmed as task/algorithm-driven, not style.
- With 2+ suspect dims, prefer CONTINUE unless conflicts resolved.

Preliminary Review usage: compare current evidence to preliminary
  hypotheses. Mark concerns as confirmed/weakened/contradicted.
  Unresolved preliminary hints are NOT evidence.

Output: {round, evidence_sufficient, assessment{
  total_dimensions_analyzed, high_confidence_dimensions[],
  resolved_conflicts[], unresolved_conflicts[],
  overall_evidence_strength}, recommendation(FINALIZE|CONTINUE),
  reasoning}
\end{lstlisting}

\subsubsection{Truth-Seeking Debate}

\begin{lstlisting}[style=promptblock,caption={Debate Participant prompt.}]
You are the {dimension} expert. Goal: FIND TRUTH, not win.
Privilege: direct CODE1/CODE2 access.

Step 1: Examine source code. Find dimension-relevant patterns,
  overlooked/overestimated evidence.
Step 2: Debate with required fields:
  claim, evidence (numbers+facts), concession,
  what_would_change_my_mind, verdict_request,
  new_code_grounded_observations, which_existing_evidence_is
    _overstated, which_existing_evidence_is_confirmed,
  updated_dimension_tendency, updated_confidence.
Principles: honest about weaknesses; proactively request deeper
  analysis when insufficient.
\end{lstlisting}

\begin{lstlisting}[style=promptblock,caption={Debate Judge prompt.}]
Debate Judge. Responsibilities:
1. Ruling: conflict resolved?
2. Tracing: dimension credibility update?
3. New Evidence: source-level insights from debate.
4. Corrections: which dimension reports need revision?

Evaluation: source consistency, preliminary review alignment,
  external consistency (one dim contradicts others?), argument
  strength (who provided more verifiable observations?).

Required tasks: determine conflict resolution, assess which side
  is more persuasive, update dimension credibility, extract new
  evidence, list remaining issues, weight adjustment recommendation,
  final judgment (same_author|different_author|uncertain),
  report patch suggestions (if a dimension's conclusion was
  overturned by new evidence, provide corrected tendency+confidence).

Output: {debate_topic, proponent, opponent, total_turns,
  resolution{conflict_resolved, explanation,
  dimension_credibility_update, weight_recommendation,
  new_evidence_from_debate, remaining_issues,
  report_patch_suggestions[]}, final_judgment,
  confidence_after_debate, recommended_next_action}
\end{lstlisting}

\subsubsection{Final Decision}

\begin{lstlisting}[style=promptblock,caption={Final Decision prompt.}]
Final Decision Judge. Directly examine CODE1/CODE2 source code,
combine all upstream evidence. NOT voting or counting anchors.

Must read: (1) CODE1+CODE2 source, (2) preliminary_review,
  (3) four dimension latest_reports, (4) latest_synthesis,
  (5) latest_debate_result, (6) latest_reflection.

Mandatory questions:
1. After re-reading code, overall gut tendency?
2. Which dims support same_author? different_author? uncertain?
3. Is preliminary confirmed/weakened/overturned?
4. Did debate bring genuinely new evidence?
5. Was reflection's advice adopted?

Hard Rules:
- No numeric anchors; uncertain != different_author.
- Overturning preliminary requires explanation.
- different_author requires >=2 moderate different dims OR 1 strong
  structural counter-evidence (confounder_risk=low) + debate.
- Mixed evidence (2 same + 2 different) => uncertain.
- Cross-lang: syntactic = weak auxiliary; same_author from
  preliminary uncertain/different requires credible non-syntactic
  primary driver (prefer lexical naming quirks).

Output: {verdict: same_author|different_author|uncertain,
  confidence, evidence_chain[4]{dimension,supports,strength,weight,
  similarity_score,confounder_risk,key_evidence},
  process_alignment{preliminary_review_status,
  dimension_alignment, debate_contribution,
  reflection_adoption}, reasoning}
\end{lstlisting}

\begin{table*}[!htbp]
\centering
\small
\vspace{-10ex}
\begin{tabularx}{\textwidth}{@{}l>{\raggedright\arraybackslash}p{4.5cm}>{\raggedright\arraybackslash}p{4.5cm}>{\raggedright\arraybackslash}p{4.5cm}@{}}
\toprule
\textbf{Dim} & \textbf{Author-Stable} & \textbf{Different-Author} & \textbf{Confounders} \\
\midrule
Layout &
  Indentation habits; brace/delimiter combos; micro-spacing; blank-line rhythm; comment placement &
  Conflicting indent units; systematic brace disagreement; different density; spacing micro-habits &
  Auto-formatters; code-size wrapping; shared boilerplate \\
Lexical &
  Identifier morphology/casing; temp-variable habits; abstract templates; operator-keyword phraseology &
  Conflicting naming discipline; different identifier lengths; different keyword phraseology &
  Problem vocabulary; language keywords; char-ngram/token overlap from templates \\
Syntactic &
  Decomposition habits; control-structure composition; structural sequencing; cross-lang prefs &
  Different decomposition; different guard/nesting style; different recursion/iteration &
  Algorithm templates (DFS/BFS/DP); parser degradation; problem complexity \\
Pattern &
  Problem-solving decomposition; complexity concentration; API/idiom selection; semantic habits &
  Different decomposition depth; different data-structure choices; different defensive habits &
  Problem requirements; ecosystem idioms; small programs with sparse evidence \\
\bottomrule
\end{tabularx}
\caption{Per-dimension focus signals for pairwise direct comparison.}
\label{tab:dimension-configs}
\end{table*}

\subsection{Pairwise Direct Comparison}
\label{sec:f-direct}

Each dimension supports a direct pairwise comparison mode on raw code. The system prompt is constructed per dimension from Table~\ref{tab:dimension-configs}.

\begin{lstlisting}[style=promptblock,caption={Pairwise direct comparison system prompt template.}]
You are a specialized {display_name} pairwise comparator for
source-code authorship attribution. You are NOT extracting features
from a single file. You are directly comparing CODE_A and CODE_B
within one stylistic dimension.

{context_line: Both snippets are in `lang`. | CODE_A in `lang_A`,
  CODE_B in `lang_B` (DIFFERENT LANGUAGES).}

Your report must separate three kinds of evidence:
1. author_stable_signals: similarities persisting across problems.
2. different_author_signals: contrasts supporting different authors.
3. neutral_or_confounding_signals: overlaps better explained by
   templates, tasks, ecosystems, or language defaults.

Prioritize: {per-dimension stable_focus items}
Different-author cues: {per-dimension different_focus items}
Actively discount: {per-dimension confounders}

Return exactly one JSON object:
{tendency, similarity_score, confidence, summary,
  author_stable_signals[], different_author_signals[],
  neutral_or_confounding_signals[], high_risk_confounders[],
  reasoning}
\end{lstlisting}

\end{document}